# A Software Development Methodology for Research and Prototyping in Financial Markets

Andrew Kumiega, Ph.D. and Ben Van Vliet
akumiega@sgfallc.com
bvanvliet@stuart.iit.edu

**Abstract**

*The objective of this paper is to develop a standardized methodology for software development in the very unique industry and culture of financial markets. The prototyping process we present allows the development team to deliver for review and comment intermediate-level models based upon clearly defined customer requirements. This spreadsheet development methodology is presented within a larger business context, that of trading system development, the subject of an upcoming book by the authors of this paper.*

**Introduction**

Few industries are driven by the need for rapid development cycles in technology as the financial markets. Over the last 20 years the rapid expansion in computing power and network sophistication and the tremendous amounts of quantitative research into the valuation and pricing of securities, commodities and derivatives, have joined technology and finance in a marriage that will never end in divorce. [Kumiega, Van Vliet, 2000] Wall Street and Technology magazine estimates that firms engaged in the financial industry spend over $42,000 per employee on technology, more than twice any other industry.

      Today, analysts and traders pour over historical price data and qualitative research looking for an edge, a new way to make money or reduce risk. Front-office financial engineers and traders employ real time data feeds and almost exclusively Microsoft's Excel spreadsheet environment for both research and implementation in a time-critical environment.

      There are several reasons for the widespread application of Excel in finance. Some of the most important are:

- The pre-built, standardized functions in Excel that can easily handle difficult and/or time consuming financial calculations, such as: DAYS360, which returns the number of days between two dates based on a 360-day year; and DURATION which returns the "duration" of a bond. (Duration is defined as the weighted average of the present value of the cash flows and is used as a measure of a bond price's response to changes in yield.)

- The embrace of Excel by most colleges and universities as the dominant tool for teaching finance. In fact, most of the new textbooks in finance include examples in both Excel and VBA code.

A Software Development Methodology for Research and Prototyping in Financial Markets: Kumeiga & Van Vliet

- The large number of commercial, market data feeds that export to Excel via Dynamic Data Exchange (DDE). Among these are firms such as Bloomberg, Spryware and Reuters.

- The ease of use that allows desk traders, who generally lack formalized education or training in programming and software engineering, to develop simple trading systems.

Given the ease of use and the urgency of results, however, individuals responsible for program development, be they either traders, financial engineers or programmers, generally produce software that suffers from several quality-related problems, among these are:

- Lack of versioning, scoping and visioning control and general disregard for the modeling life cycle proposed by Read and Batson [1999]. Financial engineers and traders almost continuously make adjustments and augmentations to their spreadsheets. This leads to multiple versions of the applications not only on an individual trader's computer, but also among different computers on the same trading desk, not to mention on different trading desks firm wide.

- Absence of any clear project framework. Therefore, it can become a very complex project to transfer these programs to a client server environment.

- Due to the difficulties in using Excel's grid environment, deficiencies in documentation of algorithms and financial calculations. This inevitably corrupts the project structure.

In some ways, Microsoft's Excel spreadsheet is the crack cocaine of the financial markets industry. It is an addictive drug and many popular financial modeling books have gotten the trading industry hooked by teaching the construction of financial applications with no consideration of planning and system design. Too often graduate level courses in financial engineering focus solely on algorithm development and not on the process of building robust trading or risk management systems. Students design Excel-based systems that should otherwise be properly implemented in object oriented programs using C# or C++.
   To be sure Excel has a well-developed network of suppliers that provide myriad add-ins, and projects in Excel are easily additive; traders and analysts simply add new spreadsheets and new calculations to existing systems. Excel's place should more often be as a prototyping tool, not an implementation tool, when constructing real-time trading systems. Prototyping implies a subsequent conversion to programming code, which often becomes a reverse engineering project by programmers that may or may not understand the spreadsheet's underlying context—the finance theory or trading strategy. Spreadsheets exist then within their business contexts and within the processes and methodologies of those contexts. The spreadsheet development methodology we present in this paper exists within the larger context of securities and derivatives trading.
   Over the last five years, our development framework has evolved from being a purely Excel-based development framework to a complete trading system development framework. In this paper we will first present this complete framework and then show how our Excel prototyping framework properly fits within it.
   Our step-by-step development methodology will control problems that arise during the development, evaluation, knowledge and personnel management, and construction of algorithmic trading systems. Furthermore, it provides a consistent framework for allocating start-up and ongoing resources for such systems. (This methodology will be presented in an upcoming book tentatively titled *Quality Financial Management*, Elsevier, 2006.) We divide the methodology into four stages which are themselves divided into four, iterative steps.

### 1.0 Overview of the Kumiega-Van Vliet Trading System Development Methodology ( **K | V** )

Trading system development is a business process which can be but is too often not managed. Too often the lack of quality execution dooms the project. What financial firms need is a systematic approach to conceive of, develop and manage new trading systems.

By their nature, all functioning trading and money management systems must manage two concurrent processes: trade selection, and portfolio/risk management. Prior to implementation, though, the process of development of such a system should follow a well-defined, well-documented flow of steps according to a development methodology.

While the peculiarities of the financial industry require a unique development paradigm, this new methodology also borrows a majority of its architecture from the traditional waterfall model [Royce, 1970], evolutionary spiral models [Boehm, 1988], and *Stage-Gate™* process models. The paradigm deviates, however, from these standardized methodologies in significant ways in order to meet the specific demands of the financial industry. By combining aspects of each of these well-known methodologies **w**e believe that our method should enable development teams to move from level 1 to level 3 on the Capability Maturity Model [SEI, 1993].

### 1.1 Waterfall Methodology

The traditional waterfall model for software development consists of four stages that will essentially map to the four stages of the K|V model--analysis, design, implementation, and ongoing system monitoring and testing.

In sum, the waterfall model forces teams to plan before building and requires a disciplined approach to development. Using the waterfall model, we should be able to avoid the pitfalls of creating systems before the plans of the project are precisely defined and approved. However, the waterfall methodology has at least two drawbacks.

The first drawback is that the waterfall tends to put too much emphasis on planning, necessitating that all details be defined up front before design and implementation can begin. As a result, it leaves no room for error and no process for handling feedback on problems that inevitably occur during development. In the fast moving financial markets, where trading opportunities come and go, the waterfall model on its own is not flexible enough to react to new information and knowledge. To overcome this shortcoming of the waterfall model, the spiral model was developed.

The second drawback is that prior to progression to each new stage, the waterfall model does not include a gate process—a management decision as to whether to continue development based upon the potential for profitable implementation. The K|V model will include gates after each stage as you will see.

### 1.2 Spiral Methodology

In the spiral methodology, a smaller amount of time is initially devoted to the four stages--research, planning, implementation, and testing--followed by several iterations or cycles over each. As the cycles progress, and the spiral gets larger, we add more detail and refinement to the project plan. At some final level of detail, each stage will be complete.

In this way, the spiral method allows for feedback as problems arise or as new discoveries are made. Problems can then be dealt with and corrected unless they are fatal. If it is determined that the trading system cannot or will not be profitable, for whatever reason, it will be discarded. So intermittent or prototype implementations can provide important feedback about the viability and profitability of a trading system.

As with the waterfall method, though, the spiral method is not without drawbacks. In the spiral methodology the cycles can grow without end and there are no constraints or deadlines to terminate the iterations. This can lead to scope creep—a loss of project focus, messy logic and unnecessary digressions—where the project plans may never contain a clear and concise system architecture. So the cycling process demands clear conditions for termination. For example, spiraling has no criteria for transition from one tool set to another say from Excel to VBA to C++.

### 1.3   *Stage-Gate ™* Methodology*

Trading system development essentially applies concepts of new product development to proprietary software design. Unlike new products though there are no external customers of the system or the software itself. Rather what we are selling are the results or the track record of the system and its interaction with the trading team. Nonetheless, we can benefit from an understanding of new product development methodologies in particular the *Stage-Gate ™* method, which was created to manage the process of new product innovation. From *Stage-Gate ™* we borrow the concept of a gate. Gates are evaluation points and the form of management meetings with team members from different functional areas. These gate meetings are where go/kill decisions are made, where weak projects are weeded out and scarce resources are reallocated towards more promising projects.

In the K|V methodology, gates act as checkpoints along the process. They check whether or not the business reason for developing the system is still valid. Our gates will each have their own set of metrics and criteria for passage. Well-organized gate meetings in the K|V model should make a decision to:

- **Go.** Go on to the next stage of the waterfall.
- **Kill.** Kill the project entirely.
- **Hold.** Hold development at the current stage for reconsideration at a future date.
- **Return.** Return to a previous stage for additional research or testing.

If the project is allowed to continue or is sent back to a previous stage the gate meeting should also outline the plan for moving through the next stage, define the deliverables expected and the criteria for evaluation at the next gate meeting. The criteria for each gate should include a check on the deliverables, minimum standards, potential for profitability, competitive advantage, technical feasibility, scalability and risk.

Essentially at each successive gate management must make a progressively stronger commitment to the project. In the end well-organized and well-run gate meetings will weed out the losers and permit worthwhile projects to continue [Cooper, 2001].

*\* Stage-Gate is a registered trademark of R.G. Cooper & Associates Consultants, Inc., a member company of the Product Development Institute. See www.prod-dev.com*

### 1.4   **Outline of the K|V Methodology**

To overcome the respective shortcomings of each of the methodologies described above, we combine them into a single paradigm for trading system development. In this expanded K|V model four stages progress in a traditional waterfall, but within each stage, four steps are connected into a spiral structure. At the completion of each stage is a gate that will allow management to kill the project, return to a previous stage or continue to the next stage of development. After completing the fourth and final stage the methodology requires that we repeat the entire four stage waterfall for continuous improvement. Here are the four stages of the K|V methodology and their respective components:

**Kumiega-Van Vliet Trading System Development Methodology ( K|V )**

**Stage I.        Research and Document Calculations**
1.      Describe Trading Idea
2.      Research Quantitative Methods
3.      Prototype in Excel
4.      Check Performance
        *Gate 1*

**Stage II.       Back Test**
1.      Gather Historical Data
2.      Develop Cleaning Algorithms
3.      Perform In Sample / Out of Sample Tests
4.      Shadow Trade and Check Performance
        *Gate 2*

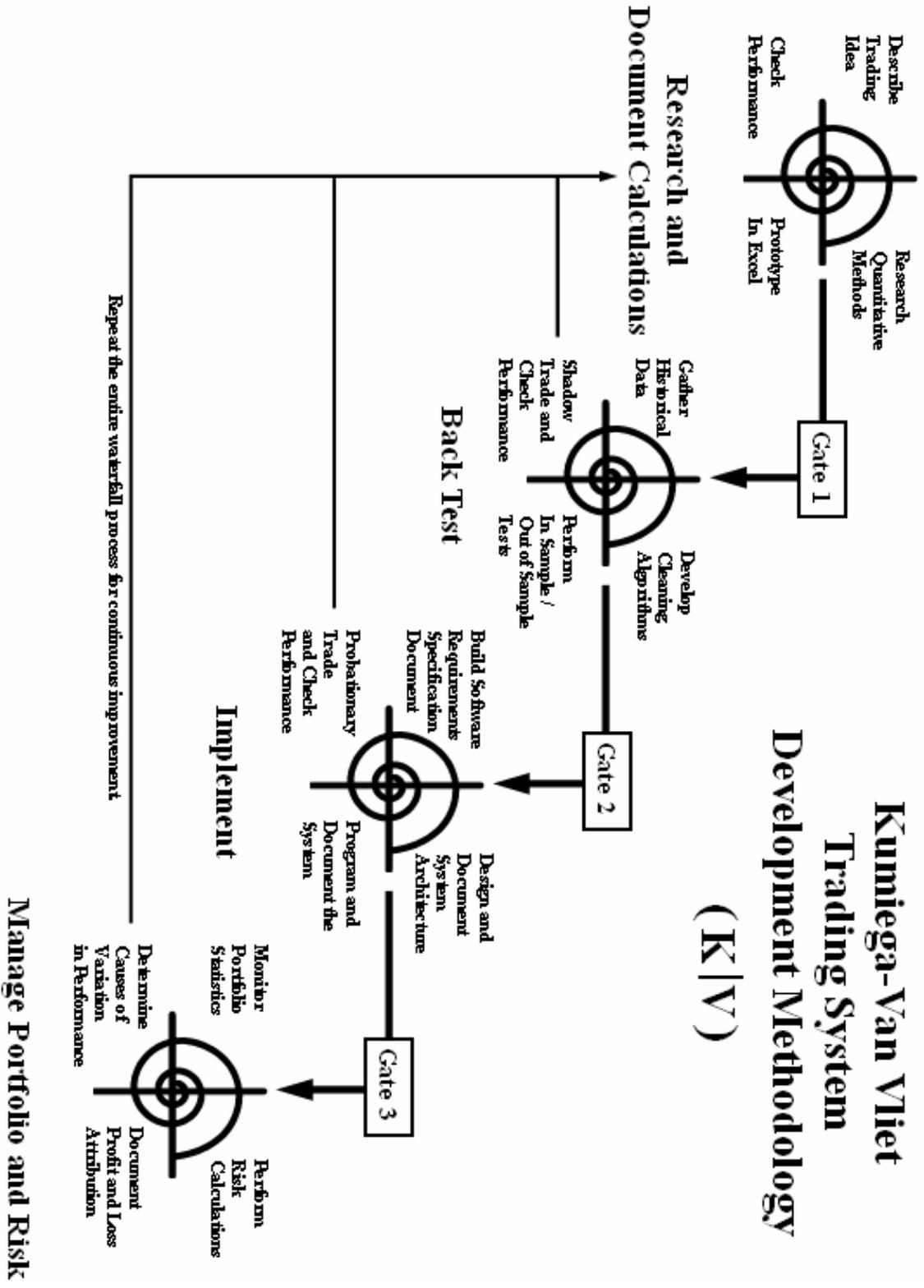

**Stage III.        Implement**
1. Build Software Requirements Specification Document
2. Design and Document System Architecture
3. Program and Document the System
4. Probationary Trade and Check Performance
   *Gate 3*

**Stage IV.        Manage Portfolio and Risk**
1. Monitor Portfolio Statistics
2. Perform Risk Calculations
3. Document Profit and Loss Attribution
4. Determine Causes of Variation in Performance

**Repeat the entire waterfall process for continuous improvement.**

As you can see, the model sets forth Excel as a prototyping tool in *Stage 1, Step 3: Prototype in Excel* (We will use the following shorthand notation to refer to the 4 stages and 16 step of the K|V Trading System Development Methodology: *K|V 1.3*). Now, we will address very specifically the problems that arise during the conversion of Excel prototypes to fully coded applications in VBA and/or subsequently into, for example, C# or C++. As we progress we will document how the methodology presented to control the conversion process integrates into the overall K|V framework.

**2.0  Prototyping in Excel**

Since the K|V methodology requires an iterative process the Excel prototyping step will be encountered several times as we spiral through stage one, as we receive inputs from K|V 1.2 and provide outputs to K|V 1.4. Quality control will start at the very first steps of implementation; we will prove and prototype small experiments and calculations.
   The final deliverable consolidates these small prototypes into a single model with one set of code. This consolidated model should be fully labeled, with code fully documented and all calculations clearly shown. Among other things a consolidated prototype will alert programmers to thread-safety issues down the road.
   With respect to the broader K|V methodology prototyping algorithms in Excel accomplishes the following goals:

- **Clear definitions of algorithms.** Building a prototype forces us to clearly define and formulate calculations. If we cannot explain the system well enough to build an Excel prototype then there is little chance that we will be able to build a working tool. Also the Excel prototypes are needed before implementation for verification.

- **Clear definition of data requirements.** Prototypes will bring into focus what raw data will be needed for implementation. Often times the required data is not publicly available or is too expensive. Further, a full prototype should include test data and user interface mock-ups for all dynamic inputs.

- **Development of white and black box results for regression testing.** These black-box results will serve as gold-standard results against which to perform regression tests in later stages of the K|V model.

- **Requirements for graphical user interfaces.** This forces discussion of about how humans will interact with the software.

- **Ability to weed out bad ideas early.** Instead of wasting weeks gathering specifications for fully coded models it is clearly advantageous to waste only an afternoon. If an idea cannot be prototyped in Excel it is not well defined and therefore must be scrapped.

It is important to note the distinction between developing trading systems and developing software. Trading systems (at least fully automated ones) are contained within software. Testing of a trading system includes testing of the software to ensure proper execution and testing of the system itself to ensure profitability. In this paper we focus on software development and software testing as a piece of the overall K|V methodology. Our standardized methodology for financial software development will ensure both a rapid development process desired by business teams and consistent quality standards desired by software professionals [Humphrey, 1995].

**3.0 Excel Prototype Loops**

Our Excel methodology begins with planning and necessitates development of a vision and scope document and prototypes in an iterative process. This document outlines the project's requirements and objectives. To understand and plan the programs to be built the project manager must understand the required behavior and performance of the customer requirements. This planning process must be based upon customer interaction. In finance the customers are the traders and investment managers who make decisions on how to allocate the firm's risk capital. The traders, quantitative analysts and developers meet to define project requirements. Prototyping then begins with requirements gathering.

**Step 1**   **Write a statement of objective with a description of the problem to be solved.**

[These activities are encompassed within K|V 1.1.]

**Step 2**   **Determine requirements and the data inputs needed for the proposed application.**

One of the major stumbling blocks is the lack of clean and timely data. Many applications built upon research analyzing historical data are never implemented due simply to the fact that real-time data is either too expensive or not available altogether. (The difference in price for real-time data to end of day data, delayed for several hours can be quite large.) Therefore, prior to starting a project the data feeds and prices should be determined first. This step greatly reduces the risk of failure.

[These activities are encompassed within K|V 1.2.]

**Step 3**   **Trader and/or quantitative analyst derive calculations.**

The next major step in development is attaining proof that the system in question can actually be solved using mathematics or at least programmable steps. Many trading ideas are multiple if...then statements that when put into equation form lack any clear objective rules of logic. On the one hand the trader believes he has a firm system, on the other the programmer sees an endless loop of "if" statements.

[These activities are encompassed within K|V 1.2.]

**Step 4**     **Determine the user interfaces using Excel as the GUI.**

Graphical users interfaces need to be laid out first since many of the systems developed depend on the traders' ability to see and react quickly. If the trader wants a live time 3-D display of risk for his 100 stocks then the particulars of how you display real-time data becomes very important. For this paper we will assume a very basic interface of inputs, outputs and a simple chart. However, that is not the norm in trading.

[These activities are encompassed within K|V 1.3.]

**Step 5**     **Prototype calculations.**

The quantitative analysts will either derive proprietary algorithms or more likely assemble research articles for equations and calculations. It will be very important to standardize the formats of the different equations as "beta" in one paper may not mean the same thing as "beta" in another. In this step we seek to break down complicated calculations into smaller ones over several cells in order to simplify the testing of intermediate results [Bewig, 2005]. Later we will consolidate models.

[These activities are encompassed within K|V 1.3.]

**Step 6**     **Test for user requirements.**

[These activities are encompassed within K|V 1.4.]

**Step 7**     **Complete a vision and scope document for the initial Excel-based product.**

Building a basic vision and scope document forces the end user to determine quickly if the product built in Excel solves the trading problem. Again the goal is to quickly stop development of projects with a low probability of success in both the short and long term.

[These activities are encompassed within K|V 1.1.]

**Step 8**     **Compare alternative quantitative methods.**

As often happens in finance, more than one different model may exist for the same purpose. For example, Black-Scholes and binomial models are both used to price options. It may be necessary to weigh the strengths and weaknesses of all alternatives.

[These activities are encompassed within K|V 1.2.]

**Step 9**     **Build a working, consolidated prototype application in Excel.**

>Here we employ the Microsoft Development Process for Rapid Application Development using the classic spiral methodology from envisioning, to planning, to developing and then to stabilization [Syngress, 1999]. If our objective is to construct a model which includes multiple small models we consolidate the prototypes into a single model. This consolidated model should be fully labeled with code fully documented and all calculations clearly shown.
>
>[These activities are encompassed within K|V 1.3.]

**Step 10  Test the application's results against the initial calculations and scope requirements.**

>[These activities are encompassed within K|V 1.4.]

**Step 11  Deliver the product to the end-user as an Excel based application.**

>The advantage of this type of approach is that it allows us to quickly deliver a prototype for evaluation by trading desk personnel with well defined features and specifications that are scaleable into say, C# or C++. The traders are then able to refine the product using the above spiral typed method of development until they either determine whether or not the product enhances the traders ability to generate revenue over a long time horizon. If the product is successful and has a high probability of long-term profitability we restart the development process with the crossover stage. This may seem counter-intuitive but frequently trading ideas produce profits only for a short amount of time, until competitors come in and remove the market inefficiency an idea was exploiting.
>
>At this point trading desk functionality testing will again invite comments from the end-users as to the effectiveness of the program. Modifications to the GUI and additional functionalities can be discussed at this time. Looping back to the initial step may be required prior to moving the project forward into the crossover stage for continued development.
>
>[Essentially this step acts as an intra-stage gate within the K|V model.]

**4.0  Crossover Loops**

Having proven a concept successful and making the determination that the application has high potential for medium-term profitability, we proceed with the crossover loops of the development process. In this stage we "crossover" from Excel's cell-based environment to Visual Basic for Applications by converting the program's functionalities into code.

**Step 1  Develop a vision and scope document for the product based on expanding the Excel vision and scope document to include Excel, Visual Basic for Applications and a database.**

>[These activities are encompassed within K|V 1.1.]

**Step 2  Convert the Excel function/cell-reference code to user-defined functions in VBA.**

>We then have the ability to continually test our calculations against Excel's black box functions.
>
>[These activities are encompassed within K|V 1.3. Notice that we have skipped K|V 1.2 under the assumption that the quantitative methods are at this point fixed. Any alterations to the quantitative methods would require a return to the Excel prototyping stage described previously.]

**Step 3  Change all of the controls to ActiveX and lock the editing of the spreadsheet.**

>[These activities are encompassed within K|V 1.3.]

**Step 4  Deliver the revised product back to the trader and allow for distribution since all of the calculations, screen layouts and charts are secured in VBA.**

>The main advantage here is that we now have a completed scope and vision document that includes:
>
>- Working calculations in VBA code
>- Completely designed graphical user interface
>- Complete data dictionary
>- Sample test cases
>- Complete buy-in from the end-users.

[These activities are encompassed within K|V 1.4.]

**Step 5  Reevaluate the usefulness of the product.**

>If the product has a long-term profitability we repeat with a true scope and vision document for development in C# or C++.
>
>[These activities are encompassed within *Gate 1* of the K|V model or would require a loop back to the Excel prototyping stage or the intra-stage gate.]

**Step 6  We can give the entire package to a programmer or outsourcing firm and receive a working product without having a programmer that is also skilled as a trader or portfolio manager.**

>The completed Excel prototype, with VBA functions, will be the foundation of the Software Requirements Specification (SRS) of K|V 3.1. The SRS will clearly document all of the trading algorithms and functionalities of the system including data dictionary, data flow maps, GUI requirements, error handling maps, and report generation maps. The SRS document will largely be based upon the Excel/VBA prototypes and descriptions created in this earlier stage of development and will allow a team of programmers to quickly build the system to the proper specifications.

**5.0 Advantages**

There are five distinct advantages to using this methodology for application development in financial markets:

- The Excel prototyping approach provides a mechanism for identifying system requirements and for buy-in by the trading team. Customers should be able to explicitly state or may even be able to demonstrate their required functionality in Excel.

- Similar to the spiral model, this methodology enables the end user and the developer to understand and react to risks at each level. The iterative framework of prototyping, testing and delivering intermediate-level working-versions for feedback reduces risks before they become problematic.

- This methodology allows for step-by-step testing against Excel's built-in functions. The testing consists of both white box and black box testing using regression.

- Outputs and layouts with upward compatibility due to the integration of Microsoft's ActiveX library.

- Time to market is greatly reduced since Excel prototype starts working in a short time. A working version of the software can be developed quickly in Excel and if need be available to customers for testing.

**6.0 Summary**

The objective of this research was to develop a standardized methodology for software development in the very unique industry and culture of financial markets. The prototyping process allows the development team to deliver for review and comment intermediate-level models based upon clearly defined customer requirements. The project's ability to add to the bottom line can be reevaluated at each step prior to the outlay of additional time and investment. Furthermore, we presented the Excel prototyping methodology within the context of a larger trading system development methodology.

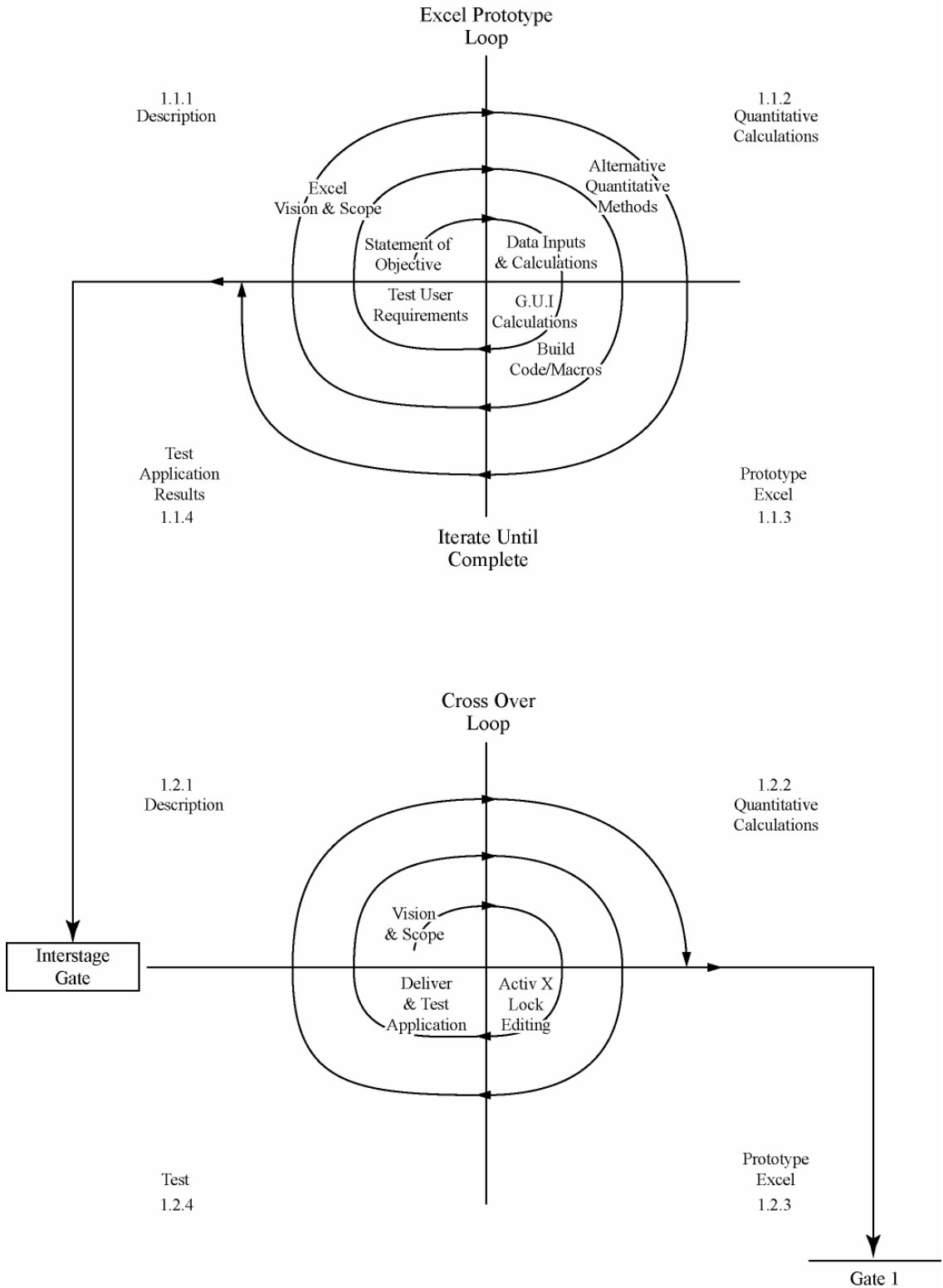

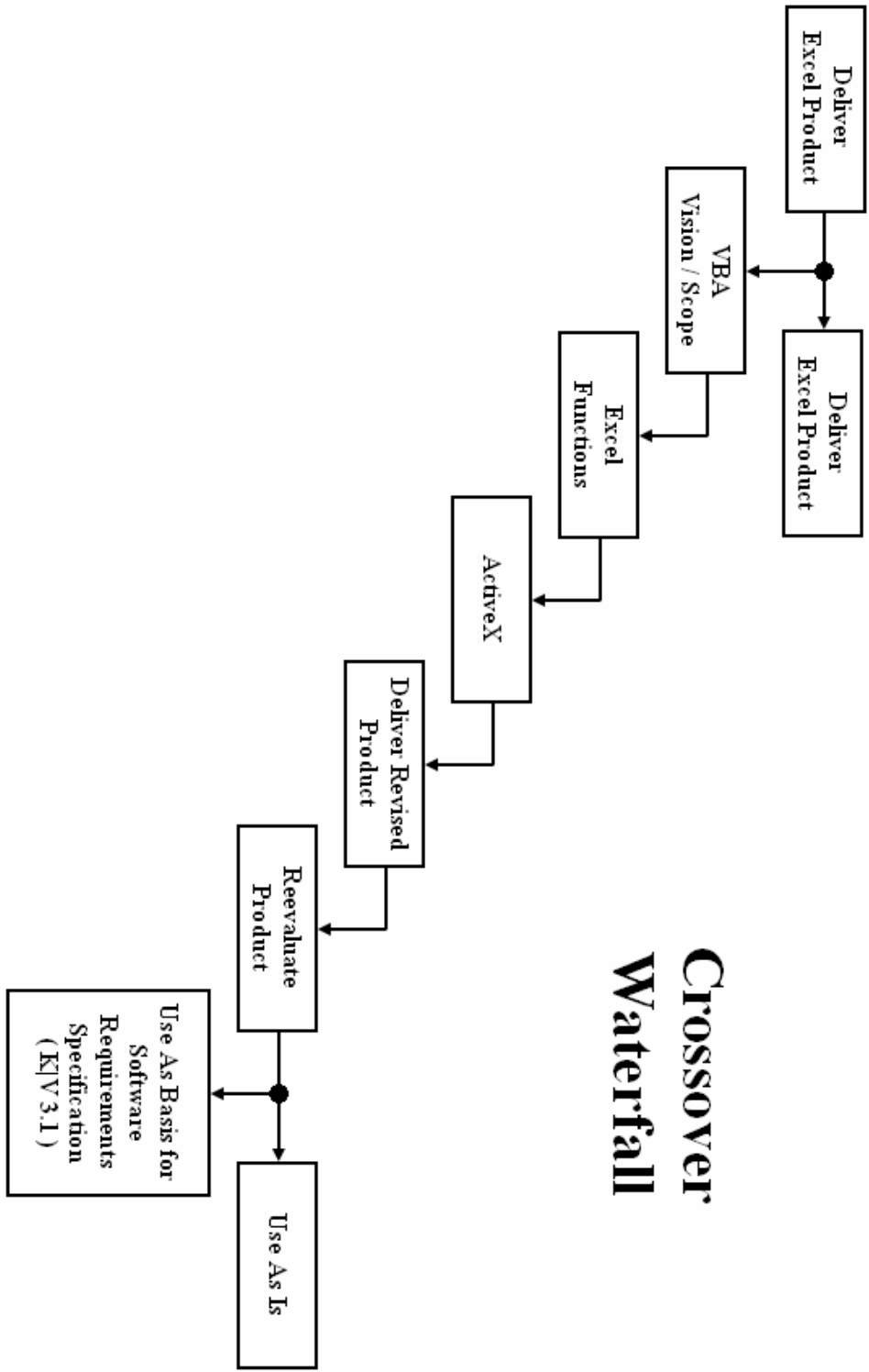

Crossover Waterfall

**7.0  Case Study**

A application of the methodology to the development of a simple software package to price call options on stocks.

**7.1  Excel Prototype Loops**

**Step 1     Write a statement of objective with a description of the problem to be solved.**

Our firm, ABC Trading, Inc., trades equity options.  After several comments from the firm's traders and subsequent interviews with them the project manager develops a vision and scope document which outlines the details of the projects.
    We are to develop a program that will allow the traders to price call options on stocks.  The program must be easy to use.  Several models exist for pricing options and consideration would otherwise be given to the strengths and weaknesses of alternative approaches.  We will develop this case study using the Black-Scholes model.  (In 1973, Fisher Black and Myron Scholes proposed a model for pricing European options on non-dividend paying stocks [Black, Scholes, 1973]).

**Step 2     Determine requirements and the data inputs needed for the proposed application.**

The inputs required to calculate the price of the call option are:

| **Ticker Symbol** | User entered |
|---|---|
| **Stock Price** | Real-time data feed (Reuters, Spryware, etc.) |
| **Expiration Date** | User entered |
| **Time till Expiration** | User entered |
| **Interest Rate** | User entered |
| **Sigma (Volatility)** | User entered |

**Step 3     Trader and/or quantitative analyst derive calculations.**

The Black-Scholes model uses the following equation to price call options:

$$C = SN(d_1) - Xe^{-rT} N(d_2)$$

where:

$$d_1 = \frac{\ln(S/X) + (r + \sigma^2/2)T}{\sigma \sqrt{T}}$$

$$d_2 = d_1 - \sigma\sqrt{T}$$

Here, C = price of the call option, S = price of the underlying stock, X = exercise price of the call, T = time till expiration of the call, r = interest rate, and σ = standard deviation of the logarithm of the stock's return, also known as the volatility.

For example, consider the following sample problem taken from John Hull's book *Options, Futures and Other Derivatives* [Hull, 1997]:

A call option has six months to expiration. The price of the underlying stock is $42, the exercise price of the option is $40, the risk-free interest rate is 10% per annum and the volatility is 20% per annum. According to the Black-Scholes model then—S = 42, X = 40, r = 0.1, T = 0.5, σ = 0.2.

$$d_1 = \frac{\ln(1.05) + 0.12(0.5)}{0.2(\sqrt{0.5})} = 0.7693$$

$$d_2 = \frac{\ln(1.05) + .08(0.5)}{.2(\sqrt{0.5})} = 0.6278$$

and,

$$Xe^{-rT} = 40e^{-0.05} = 38.049$$

The value of the European call, then, is

$$C = 42N(0.7693) - 38.049N(0.6278)$$

Using a polynomial approximation,

$$N(0.7693) = 0.7791$$
$$N(0.6278) = 0.7349$$

So that,

$$C = 4.76$$

The price of the call option then is $4.76.

**Step 4    Determine the user interfaces using Excel as the GUI.**

For this paper we will assume a very basic interface of inputs, outputs and a simple chart. Again this is not the norm in trading.

**Step 5    Prototype calculations.**

|     | A          | B       | C                                        |
|-----|------------|---------|------------------------------------------|
| 1   | Black-Scholes Option Pricing Engine | | |
| 2   |            |         |                                          |
| 3   | Ticker     | IBM     | = User entered                           |
| 4   |            |         |                                          |
| 5   | S          | 42      | = Realtime datafeed                      |
| 6   | X          | 40      | = User entered                           |
| 7   | T          | 0.5     | = User entered                           |
| 8   | r          | 10.00%  | = User entered                           |
| 9   | sigma      | 20%     | = User entered                           |
| 10  |            |         |                                          |
| 11  | $d_1$      | 0.7693  | =(LN(B5/B6)+(B8+B9^2/2)*B7)/(B9*SQRT(B7)) |
| 12  | $d_2$      | 0.6278  | =B11-B9*SQRT(B7)                         |
| 13  |            |         |                                          |
| 14  | $N(d_1)$   | 0.7791  | =NORMSDIST(B11)                          |
| 15  | $N(d_2)$   | 0.7349  | =NORMSDIST(B12)                          |
| 16  |            |         |                                          |
| 17  | Call Price | 4.76    | =B5*B14-B6*EXP(-B8*B7)*B15               |

And, we have a graph in Excel, which shows the payoff diagram of the option at expiration.

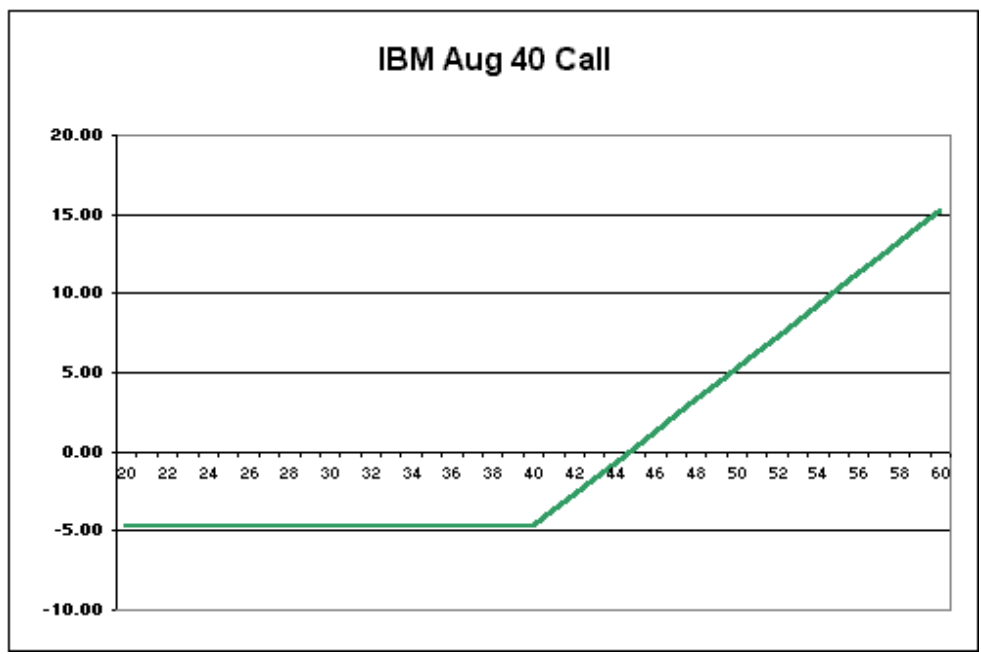

**Step 6   Test for User Requirements.**

- Screen layouts with simulated data and video tapes.
- Rough P/L calculated with live playback / regression testing.

- Mock trading using real time data.
- Live time testing.

**Step 7    Complete a vision and scope document for the initial Excel-based product.**

**Step 8    Compare alternative quantitative methods.**

Again for simplicity we assume that alternative methods for pricing options have been investigated and are now rejected in favor of the Black-Scholes model.

**Step 9    Build a working, consolidated prototype application in Excel.**

Had our customers requested a product that included more than one quantitative model we would consolidate prototypes at this stage.

**Step 10   Test the application's results against the initial calculations and scope requirements.**

As we can see here our program is running properly.  The testing follows the standard quality test procedures for both process and software.

- Regression testing for calculations.  A data tape of input is captured from a standard data provider.
- The tape is replayed for applications.  Prices can be calculated and regression tested against the theoretical.
- P/L of the system and ability of the trader to interface with the system can be captured using databases and video.
- Standard statistical analysis of the P/L can be compared with the projected P/L for a sampling period using automated trade execution or specific time of day periods with the human operator.

**Step 11   Deliver the product to the end-user as an Excel based application.**

**7.2  Crossover Loops**

Let's assume that our new program is very successful and is popular among the traders.  We now move to the crossover loops to further develop our program in code.

**Step 1    Develop a vision and scope document for the product based on expanding the Excel vision and scope document to include Excel, Visual Basic and a database.**

**Step 2    Convert the Excel function/cell-reference code to user-defined functions in VBA.**

```vb
Function BlackScholesCall(Stock, Exercise, Time, InterestRate, Sigma)

    Dim d1 As Double, d2 As Double, Nd1 As Double, Nd2 As Double

    ' Calculate d1 and d2

        d1 = (Log(Stock / Exercise) + (InterestRate + (Sigma ^ 2) / 2) * Time) / (Sigma * Sqr(Time))
        d2 = d1 - Sigma * Sqr(Time)

    ' Calculate N(d1) and N(d2)

        Nd1 = NormalCDF(d1)
        Nd2 = NormalCDF(d2)

    ' Calculate the price of the call

        BlackScholesCall = Stock * Nd1 - Exercise * Exp(-InterestRate * Time) * Nd2

End Function

Function NormalCDF(X)

    ' Calculate the cumulative probability distribution function for standard normal at X

    NormalCDF = Application.NormSDist(X)

End Function
```

We can now verify that our user-defined functions run properly by comparing our solutions against the black-box Excel functions.

| 17 | Call Price | 4.76 | =B5*B14-B6*EXP(-B8*B7)*B15 |
|----|-----------|------|----------------------------|
| 18 |           |      |                            |
| 19 | Call Price | 4.76 | =BlackScholesCall(B5,B6,B7,B8,B9) |

Next, we can escape our dependence on Excel's NORMSDIST function by approximating the standard normal cumulative distribution function in the following way.

```vb
Function NormalCDF(X)

    ' Calculate the cumulative probability distribution function for standard normal at X

    Dim a As Double, b As Double, c As Double, d As Double, prob As Double

    a = 0.4361836
    b = -0.1201676
    c = 0.937298
    d = 1 / (1 + 0.33267 * Abs(X))

    prob = 1 - 1 / Sqr(2 * 3.1415926) * Exp(-0.5 * X * X) * (a * d + b * d * d + c * d * d * d)

    If X < 0 Then prob = 1 - prob

    NormalCDF = prob

End Function
```

Again we are able to verify that our new NormalPDF user-defined function runs properly by comparing our solution against the black-box Excel functions.

| 17 | Call Price | 4.76 | =B5*B14-B6*EXP(-B8*B7)*B15 | |
| 18 | | | | |
| 19 | Call Price | 4.76 | =BlackScholesCall(B5,B6,B7,B8,B9) | |

**Step 3**     **Change all of the controls to ActiveX and lock the editing of the spreadsheet.**

Now that we are satisfied that our calculations are running correctly we can set about creating a graphical user interface using ActiveX objects.

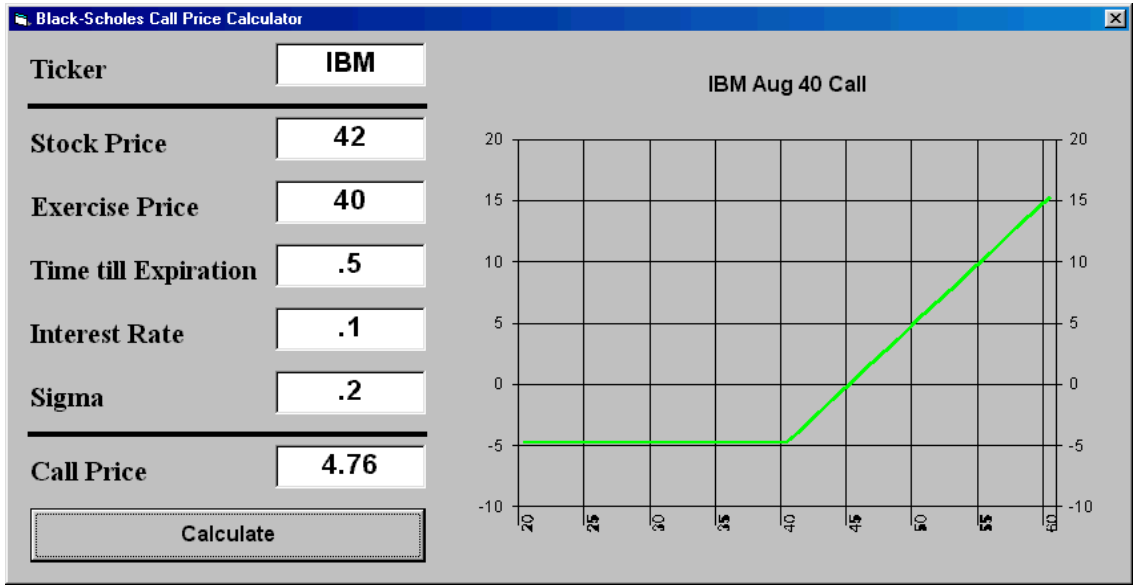

Once again we can verify our answers back against Excel's solutions.

**Step 4**     **Deliver the revised product back to the trader and allow for distribution since all of the calculations, screen layouts and charts are secured in VBA.**

Now we will have a completed scope and vision document that includes working calculations of our option pricing model in Visual Basic code. We also have a completely designed graphical user interface. We will also have a complete data dictionary, sample test cases, and complete buy-in from the traders.

**Step 5**     **Reevaluate the usefulness of the product.**

If the product adds to the long-term profitability of the traders' activities we restart the process with a true scope and vision document for development in Visual Basic or C++.

**Step 6**     **We can give the entire package to a programmer or outsourcing firm and receive a working product without having a programmer that is also skilled as a trader or portfolio manager.**

## 6. REFERENCES

Kumiega, Andrew and B. Van Vliet. 2000. "Obsolescence of the Naked Trader." *Journal of Global Financial Markets* 3 (Winter): 21-23

Black, Fisher, and M. Scholes. 1973. "The Pricing of Options and Corporate Liabilities." *Journal of Political Economy* 81 (May-June): 637-654

Hull, John C., 1997. *Options, Futures, and Other Derivatives*. Englewood Cliff, NJ: Prentice Hall, p. 244.

Boehm, Barry W., "A Spiral Model of Software Development and Enhancement," *Computer*, Volume 21, Number 5, (May 1988), 61-72.

Royce, Winston W., "Managing the Development of Large Software Systems," *Proceedings of IEEE WESCON* (August 1970), 1-9.

Humphrey, Watts S., 1995. *A Discipline for Software Engineering,* Addison-Wesley Publishing Company, Reading, Massachusetts.

*Capability Maturity Model for Software, Version 1.1*, Document No. CMU/SEI-93-TR-24, ESC-TR-93-177 (Carnegie Mellon University Software Engineering Institute, Pittsburgh, Pennsylvania, 1993)

Syngress Media*, MCSD Analyzing Requirements Study Guide (Exam 70-100)*, 1999. Berkeley, CA. Osborne McGraw-Hill, p. 55.

Cooper, Robert G. 2001. *Winning at New Products*. Basic Books. Cambridge, MA.

Read, Nick and J. Batson. April, 1999. "Spreadsheet Modelling Best Practice," Business Dynamics, and the Institute of Chartered Accountants for England and Wales.

Kumiega, Andrew and Ben Van Vliet. 2001. October 23. "A Software Development Methodology for Financial Markets." Paper presented at the *11th International Conference on Software Quality*. Pittsburgh, PA.

Bewig, Philip L. July, 2005. "How do you know your spreadsheet is right?"

Blank page